# Can crystal symmetry reshape ENZ photonics?: Opinion

MÁRIO G. SILVEIRINHA,[1*]

[1] *University of Lisbon – Instituto Superior Técnico and Instituto de Telecomunicações, Avenida Rovisco Pais 1, 1049-001 Lisbon, Portugal*

**mario.silveirinha@tecnico.ulisboa.pt*

**Abstract:** Over the last decade, epsilon-near-zero (ENZ) photonics has been driven by the search for lower losses and stronger nonlinear responses. Here, I ask a different question: can crystal symmetry also be used to shape the ENZ response? I focus on low-symmetry conductors, where the geometry of the electronic states can produce electric currents that are not present in ordinary Drude materials. These currents may enable polarization-dependent gain, nonreciprocal effects, and ultrafast nonlinearities controlled by symmetry and an external bias. I suggest that such materials may be useful for active and time-varying nanophotonics.

## 1. Introduction

Materials with a permittivity close to zero have attracted considerable attention [1-2]. Near the ENZ point, small changes in the material response can produce large optical effects. This has enabled strong nonlinearities, ultrafast optical modulation, and, more recently, time-varying photonic responses based on transparent conducting oxides such as ITO [3-8]. Most of this work has focused on the strength and speed of the nonlinear response. However, other aspects of the ENZ response have received less attention, including the role of crystal symmetry in the nonlinear response near the ENZ point.

Low-symmetry conductors, such as certain Weyl semimetals and transition-metal dichalcogenides, are particularly interesting in this context. In the last decade, these materials have attracted considerable interest in condensed matter physics, following the prediction and observation of the nonlinear Hall effect [9]. Unlike the ordinary Hall effect, this response does not require broken time-reversal symmetry. Instead, it can occur in conductors that preserve time reversal but lack inversion symmetry. The microscopic origin of the effect is related to the quantum geometry of the electronic states. Specifically, the Berry curvature near the Fermi surface can have a nonzero dipolar moment, usually referred to as the Berry curvature dipole [9, 10].

The Berry curvature dipole has been mostly discussed in the context of electron transport. However, its significance goes beyond low-frequency nonlinear Hall physics. In a conductor, the same quantum mechanism contributes to the electric current and can tailor the material response seen by electromagnetic waves [11-14]. To see this, it is useful to consider a minimal phenomenological model [15].

## 2. Minimal electrodynamic model

In an ordinary conductor, the electrodynamics of free carriers can be described, at the simplest level, by a Drude equation for the electron momentum,

$$\frac{d\mathbf{p}}{dt} + \Gamma \mathbf{p} = -e\mathbf{E} . \quad (1)$$

with $e$ the elementary charge. The electric current $\mathbf{j} = -en\mathbf{v}$ follows from the relation between momentum and velocity, $\mathbf{v} = \mathbf{p} / m^*$, so that the material response is controlled by the effective mass $m^*$, the carrier density $n$, and the scattering rate $\Gamma$. This is the standard picture behind the Drude permittivity and, in particular, behind the ENZ response of transparent conducting oxides.

In low-symmetry conductors this picture is incomplete. Bloch electrons may acquire an additional velocity contribution, known as the anomalous velocity, which is governed not by the band dispersion alone but by the Berry curvature of the electronic states [10]. After averaging over the states near the Fermi surface, its effective contribution can be written phenomenologically as a term of the form $\mathbf{v}_{\text{an}} \sim (\mathbf{p}\cdot\mathbf{D})\times\mathbf{E}$, where $\mathbf{D}$ denotes the dimensionless Berry-dipole tensor [9, 13-15]. The precise tensor structure is fixed by the point group of the crystal. Thus, in addition to the usual effective-mass term, the carrier velocity contains a quantum-geometric contribution that is controlled by symmetry, so that the current takes the form [15]:

$$\mathbf{j} = -ne\frac{\mathbf{p}}{m^*} + \frac{e^2}{\hbar^2}(\mathbf{p}\cdot\mathbf{D})\times\mathbf{E}\,. \tag{2}$$

The second contribution is nonlinear. It depends on the product of the carrier momentum and the electric field. As a result, when the low-symmetry conductor is driven by an optical or static electric field, the current density acquires an additional $\mathbf{D}$-dependent contribution, which is absent in centrosymmetric Drude metals. This is the mechanism by which crystal symmetry enters the plasmonic response.

Low-symmetry plasmonic materials have a key characteristic, also present in conventional ENZ conductors: free carriers can move over a much larger distance within one optical cycle than bound electrons in a dielectric under comparable conditions. This makes very large nonlinearities possible, especially when the electronic response is combined with plasmonic or ENZ field enhancement. Experiments have already reported giant second-order infrared responses in low-symmetry Weyl and Berry-dipole materials, including TaAs-family semimetals and two-dimensional tellurium [16-18].

There is, however, an important difference from the nonlinearities in conducting-oxide ENZ materials. Because it originates from a non-centrosymmetric electronic structure, the Berry-dipole contribution is of $\chi^{(2)}$-type, whereas conventional ENZ nonlinearities are most often described as effective $\chi^{(3)}$-responses. From Eq. (2), a bias field $E_{\text{bias}}$ gives a tensorial effective permittivity change of order:

$$\Delta\varepsilon_{\text{BD}} \sim \varepsilon_0\chi^{(2)}_{\text{BD}}E_{\text{bias}}\,, \qquad \chi^{(2)}_{\text{BD}} \sim D\frac{1}{\omega(\omega+i\Gamma)}\frac{e^3}{\varepsilon_0\hbar^2}\,, \tag{3}$$

where $\chi^{(2)}_{\text{BD}}$ is the equivalent second-order coefficient and $D$ is the relevant element of the Berry-dipole tensor. The factor $1/[\omega(\omega+i\Gamma)]$ relates to the Drude dynamics of the carriers. At lower frequencies, a free electron can swing over a larger distance during one optical cycle, which enhances the nonlinear response. Reported first-principles values of $D$ span a broad range, from $D \sim 10^{-2}$ to values well above unity, reaching $D \sim 10-20$ in some low-symmetry semimetals, such as NbAs and NbP [12, 19, 20].

To illustrate the scale, consider TaAs. First-principles calculations give $D \approx 0.39$ [19]. Using Eq. (3) at $\lambda = 800nm$ gives $\chi^{(2)}_{\text{BD}} \sim 3000$ pm/V, close to the giant second-harmonic coefficient reported experimentally in TaAs at the same wavelength [16]. This experiment was not done at the linear-response ENZ point. The free-carrier plasma crossing of TaAs is estimated to occur at much longer wavelengths, $\lambda_{\text{ENZ}} \sim 15-30\ \mu$m, depending on polarization [21].

To give an idea of the magnitude, Fig. 1 plots Eq. (3) in the weak-loss limit for representative Berry-dipole values. The estimate reaches conventional nonlinear-crystal values already for modest *D*, and grows rapidly at longer wavelengths. The dashed line marks 380pm/V for GaAs near a favorable wavelength of 800nm. The response can be even larger near electronic resonances not described by Eq. (3). For example, Ref. [17] reported a resonance-enhanced nonlinear response in TaAs near 1.77μm as large as $\chi^{(2)} \sim 10^5$ pm/V .

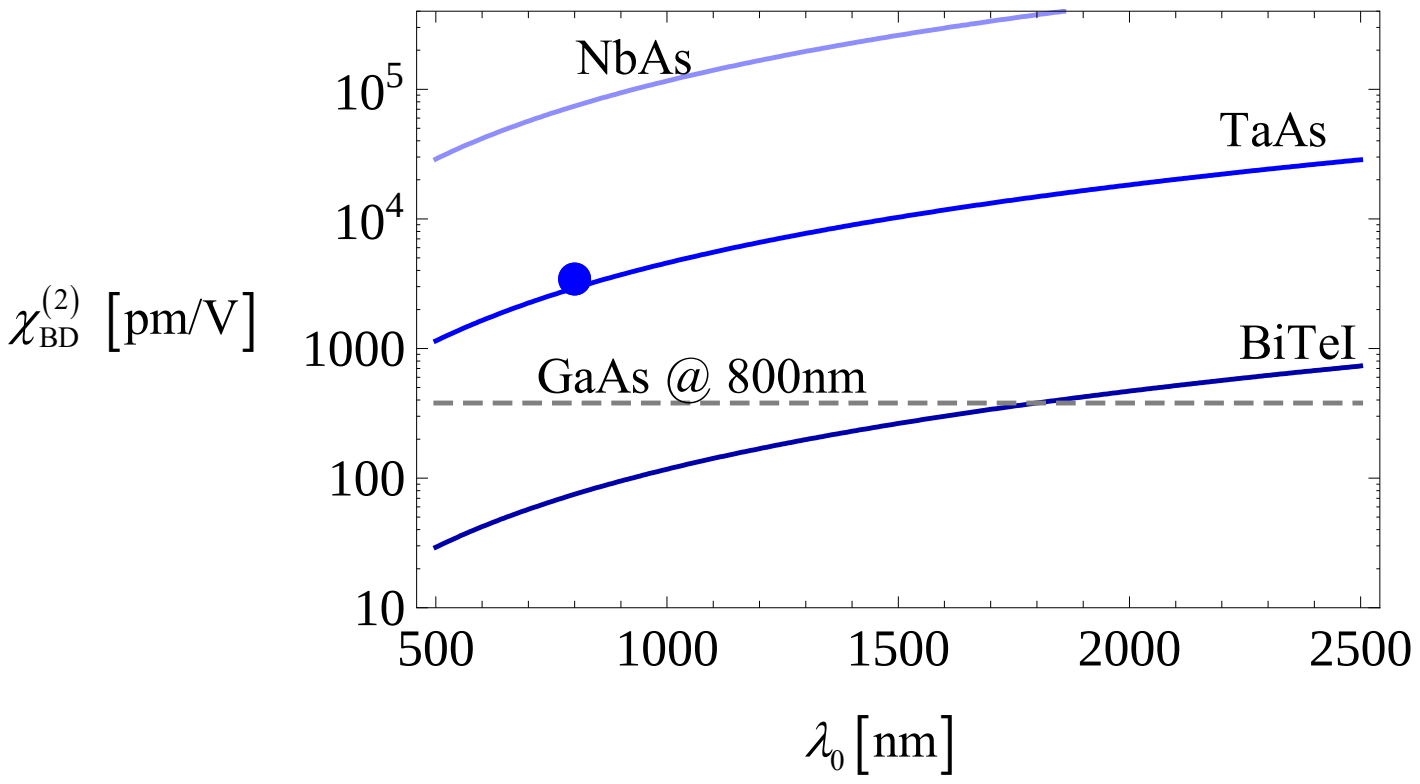

Fig. 1 Equivalent $\chi^{(2)}_{\mathrm{BD}}$ estimated from Eq. (3) in the $\omega >> \Gamma$ limit, for *D*=0.01, 0.39, and 9.88, representative of values reported for BiTeI, TaAs, and NbAs [19, 20]. The dashed line marks 380pm/V for GaAs near a favorable wavelength (800nm). The blue dot is the experimental result of Ref. [16], far from the resonance-enhanced nonlinearity at 1770nm (not shown).

## 3. Electro-optic response and chiral gain

As discussed above, the Berry-dipole produces a nonlinear current. A static bias makes this nonlinearity directly useful for linear optics. If the electric field is written as the sum of a dc component and a small optical field, $E = E_{\mathrm{bias}} + E_{\omega} e^{-i\omega t}$ , the $\chi^{(2)}$-type Berry-dipole current contains terms that are linear in $E_{\omega}$ and proportional to $E_{\mathrm{bias}}$ [13, 14]. In this way, a static bias converts the Berry-dipole nonlinearity into an electro-optic contribution to the optical conductivity. In particular, a biased Berry-dipole conductor can become nonreciprocal even without magnetic order [13, 14]. The nonreciprocal response is made possible by the bias-driven dc current, which breaks time-reversal symmetry. This effect was experimentally demonstrated in tellurium almost 50 years ago. Specifically, Vorob'ev and co-workers showed that an electric current can induce a Faraday-like rotation of mid-infrared light in Te [11].

The same electro-optic contribution can also be non-Hermitian: depending on the polarization of the wave, the optical field can suffer additional attenuation or experience amplification due to interactions with the biased material. The amplification mechanism is closely related to that underlying the operation of transistor-based microwave amplifiers [13, 14]. The wave polarizations that experience gain or loss are controlled by the crystal symmetry [22]. This is the origin of chiral gain: for some point groups, the gain and dissipative responses are governed by the handedness of the optical field, so that one circular polarization is amplified whereas the opposite one is attenuated [13, 14, 22, 23].

The chiral-gain effect can be particularly interesting for plasmonics. Surface plasmons have a transverse spin locked to the propagation direction. Since, in a biased Berry-dipole conductor, gain and loss can be tied to the spin angular momentum (i.e., the handedness of the wave), the electro-optic response can lead to gain-momentum locking: the amplified surface plasmon is selected by its direction of propagation [24, 25].

Low and collaborators have recently developed a more general theory of metallic electro-optic effects in biased quantum materials, including mechanisms beyond the Berry-dipole contribution [26-27]. They also investigated the optical gain produced by these effects. The Berry-dipole mechanism has also been considered in photovoltaics and energy harvesting, where nonlinear Hall rectification can convert optical fields into dc power [28, 29].

## 4. Outlook

The question raised here is whether low-symmetry conductors can extend ENZ photonics. Similar to conventional ENZ materials, they combine free-carrier motion and plasmonic confinement. The difference is that they can have a $\chi^{(2)}$-type nonlinear response controlled by crystal symmetry and by the applied bias.

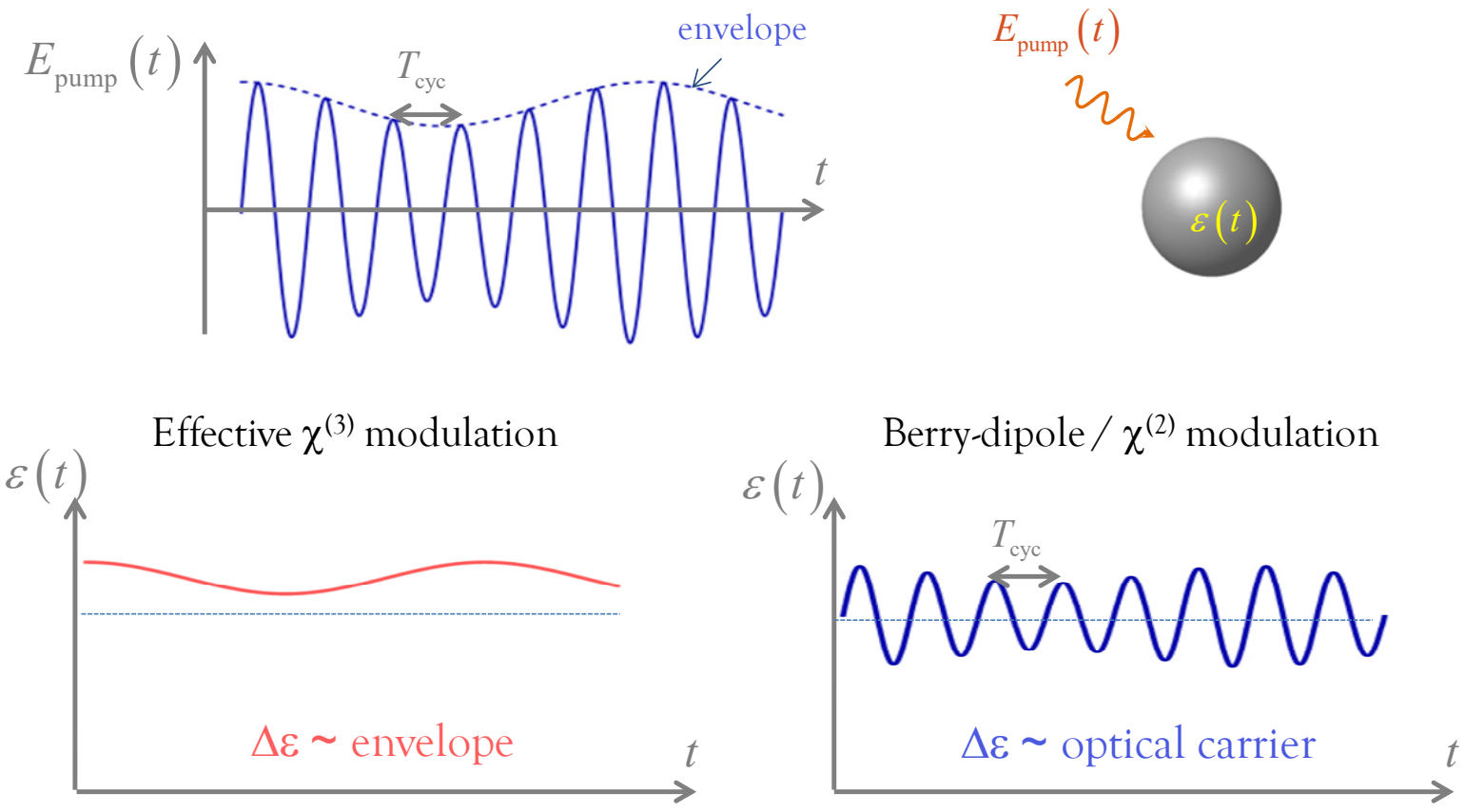

Fig. 2 Schematic comparison between effective $\chi^{(3)}$-type and $\chi^{(2)}$-type modulations. In conventional ENZ platforms, the dominant nonlinear response is often associated with the optical intensity and therefore follows the envelope of the excitation. A $\chi^{(2)}$-type response can instead follow the optical field itself, enabling modulation at the optical-cycle level ( $T_{cyc}$ ).

This may be especially relevant for time-varying photonics [30, 31]. In conventional conducting-oxide ENZ platforms, the dominant nonlinear response is often described as an effective $\chi^{(3)}$ process. This mechanism naturally leads to envelope modulation. As illustrated in Fig. 2, a $\chi^{(2)}$-type Berry-dipole response is qualitatively different. With an optical pump acting as the bias field, a $\chi^{(2)}$ nonlinearity mixes fields directly at the optical-cycle level [32-34]. Thus, it may provide a route to ultrafast temporal modulation, parametric amplification, and spacetime-crystal behavior generated by the material response itself [34].

Only experiments can determine how far this idea can be taken. Berry-dipole ENZ materials will not avoid the usual difficulties of plasmonics. Loss and heating will matter, as they do for ITO and other conducting oxides. The relevant question is whether low-symmetry conductors can combine a useful ENZ frequency, acceptable dissipation, and a strong $\chi^{(2)}$-type response. Existing nonlinear Hall and second-harmonic experiments already show that the underlying quantum-geometric nonlinearities can be large [16-18].

Crystal symmetry has been central in nonlinear optics for decades, but it has played a limited role in the ENZ discussion, where strength and speed have dominated. Berry-dipole conductors suggest that symmetry may also shape the ENZ response itself. It may therefore be worth experimentally testing the potential of some of these low-symmetry conductors for ENZ nanophotonics.

## Acknowledgments

This work was partially funded by the Simons Foundation Award SFI-MPS-EWP-00008530-10 and by national funds through FCT – Fundação para a Ciência e a Tecnologia, I.P., and, when eligible, co-funded by EU funds under project/support UID/50008/2025 – Instituto de Telecomunicações, with DOI identifier <https://doi.org/10.54499/UID/50008/2025>.

## Disclosures

The author declares no conflicts of interest.

## Data availability

Data sharing is not applicable to this article as no new experimental data were generated.